# Mixed-gap vector solitons in PT-symmetric lattices with saturable nonlinearity

**Lei Li**[1,3], **Xiaoguang Yu**[1,3], **Xing Zhu**[2], **Baiyuan Yang**[1,3]**, Qianglin Hu**[1,3] **and Xiaobing Luo**[1,3,4]

[1] Department of Physics, Jinggangshan University, Ji'an 343009, China

[2] Department of Physics, Guangdong University of Education, Guangzhou 510303, China

[3] Institute of Atomic and Molecular Physics & Functional Materials, Jinggangshan University, Ji'an 343009, China

[4] Author to whom any correspondence should be addressed

E-mail: xiaobingluo2013@aliyun.com

**Abstract**

We have investigated mixed-gap vector solitons involving incoherently coupled fundamental and dipole components in a parity-time (PT) symmetric lattice with saturable nonlinearity. For the focusing case, vector solitons emerge from the semi-infinite and the first finite gaps, while for the defocusing case, vector solitons emerge from the first finite and the second finite gaps. For both cases, we find that stronger saturable nonlinearity is relative to sharper increase/decrease of soliton power with propagation constant and to narrower existence domain of vector solitons. This finding is helpful for realizing high-power solitons with limited range of propagation constant. Additionally, our numerical calculations show that increasing the weight of dipole component results in destabilization of vector solitons, while stronger saturable nonlinearity to certain extent suppresses the instability of vector solitons.

# 1. Introduction

Vector solitons are multi-component self trapped states when several fields interact nonlinearly [1]. In the last decades, there have been extensive studies on vector spatial solitons formed by the incoherent superposition of two light beams [1–10]. Mutually incoherent two-component vector spatial solitons were first theoretically proposed in photorefractive crystals [2] and experimentally demonstrated [3]. Xu *et al* [4] investigated the stability of vector solitons in nonlocal nonlinear media. Yang *et al* [5] studied stable vortex and dipole vector solitons in saturable nonlinear media. Besides, there have been many reports on vector solitons created in lattice potentials [6–10]. It has been demonstrated by Kartashov *et al* [6] that in the vector case unstable families of scalar lattice solitons might be stabilized through the nonlinear cross-phase modulation (XPM) effect. Mixed-gap vector solitons whose components emerge from different gaps of the Bloch spectrum of the periodic lattice had also been introduced [7-9]. In experiments, the formation of fundamental and dipolelike vector solitons in an optically induced two dimensional photonic lattice had been demonstrated [10].

On the other hand, a pioneering work introduced by Bender and his coworkers in 1998 revealed that non-Hermitian Hamiltonians might have real spectra, on condition that the associated complex potential $R(x)$ satisfies the necessary condition $R(x) = R^*(-x)$, i.e., the real and imaginary parts of the complex potential should be even and odd functions of position, respectively. In optics, this situation can be readily realized by involving symmetric index guiding and an anti-symmetric gain/loss profile [11]. Such PT optical media have been created experimentally [12,13]. Optical solitons in Kerr media modulated by PT-symmetric localized and periodic potentials were first analyzed in 2008 [14], which shows that stable PT solitons can exist provided that the loss modulation parameter is below breaking point. Since then, spatial solitons have been studied in various types of PT-symmetric optical systems, including Bessel potential [15], Rosen-Morse potential [16], defect lattice [17], mixed linear-nonlinear lattices [18], superlattice [19], nonlocal media [20], quadratic nonlinear media [21], saturable nonlinear media [22,23]. Recent experiment observation has also confirmed optical solitons in PT-symmetric lattices [24]. Vector solitons in Kerr nonlinear media modulated by a PT linear lattice and by a PT mixed linear-nonlinear lattice have been studied by Kartashov *et al* [25] and Zhu *et al* [26] successively. Very recently there is the report about nonlocal incoherent vector PT spatial solitons [27]. Additionally, saturable nonlinear media are known to be promising for the control of solitons [22, 23]. However, two questions important in the field have so far not been studied, that is, the existence and the stability of vector solitons in PT-symmetric lattices with saturable nonlinearity.

In this paper, we investigate two-component vector solitons in a 1D PT-symmetric lattice in the presence of saturable nonlinearity. We demonstrate, for the first time to our knowledge, that mixed gap PT vector solitons composed of incoherently coupled fundamental and dipole components can be formed in both the focusing and defocusing saturable nonlinear media. For focusing media, the fundamental component belongs to the semi-infinite gap while the dipole component to the first gap. For defocusing media, the fundamental component belongs to the first gap while the dipole component to the second gap. Besides, saturable nonlinearity is usable to suppress the instability of such vector soliton at a certain level and saturation degree has significant influence on the power and the existence range of vector solitons. However, relatively high power of the dipole component may lead to destabilization of such vector solitons. These results may be used to realize stable "high-power" solitons within a narrow range of propagation constant.

**2. Modeling existence and stability of Vector Solitons**

The propagation of two incoherently interacting beams in 1D PT-symmetric lattices with saturable nonlinearity can be described by the normalized coupled nonlinear Schrödinger equations in a generic form [5,25]

$$i\frac{\partial U_{1,2}}{\partial z} + \frac{1}{2}\frac{\partial^2 U_{1,2}}{\partial x^2} + [V(x)+iW(x)]U_{1,2} + \sigma\frac{(|U_1|^2+|U_2|^2)U_{1,2}}{1+s(|U_1|^2+|U_2|^2)} = 0, \quad (1)$$

where $U_1$ and $U_2$ are the normalized field amplitudes of two beams, $z$ and $x$ stand for the normalized longitudinal and transverse coordinates respectively; $V(x)$ and $W(x)$ are the real and imaginary parts of the PT-symmetric potential; The $s$ is saturation parameter standing for the degree of saturable nonlinearity. The coefficient $\sigma = \pm 1$ represents "focusing" and "defocusing" nonlinearity, respectively.

In this paper, we consider the PT-symmetric lattice

$$V(x) = V_0\cos(2x), \ W(x) = W_0\sin(2x), \quad (2)$$

where $V_0$ and $W_0$ represent the modulation depth of the real and imaginary parts of the PT symmetric lattice. In this paper, we set $V_0 = 3$, and $W_0 = 0.75$. The real and imaginary parts of the PT-symmetric optical potential are plotted in figure 1(a) by blue solid and red dotted curves. The Bloch band-gap structure is also calculated, as seen in figure 1(b), with gaps from top to bottom called sequentially as the semi-infinite gap ($\mu \geqslant 1.351$), first finite gap ($1.324 \geqslant \mu \geqslant -1.261$), second finite gap ($-1.66 \geqslant \mu \geqslant -2.982$), and so on herein.

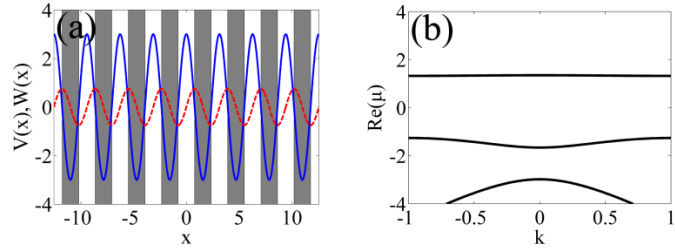

**Figure 1.** (a) The schematic of the PT-symmetric lattice. (Blue solid and red dashed lines represent the real and imaginary parts of the PT-symmetric potential, respectively. The white stripes represent high-potential regions of the real part of the PT potential, while the gray stripes represent low-potential regions). (b) The Bloch band-gap structure of the system.

We search the stationary soliton solutions of (1) with the form of

$$U_{1,2}(x,z) = f_{1,2}(x)\exp(i\mu_{1,2}z), \qquad (3)$$

where $\mu_{1,2}$ are nonlinear propagation constants of two components and $f_{1,2}$ the complex functions.

Substituting (3) into (1) leads to

$$\frac{1}{2}\frac{\partial^2 f_{1,2}}{\partial x^2} + [V(x)+iW(x)]f_{1,2} + \sigma\frac{(|f_1|^2+|f_2|^2)f_{1,2}}{1+s(|f_1|^2+|f_2|^2)} - \mu f_{1,2} = 0. \qquad (4)$$

Soliton solutions $f_{1,2}$ can be explored by solving (4) numerically [28], and the total ($P$) and partial ($P_{1,2}$) power of the vector soliton calculated by $P = \int_{-\infty}^{+\infty}(|f_1|^2+|f_2|^2)dx$, $P_1 = \int_{-\infty}^{+\infty}|f_1|^2 dx$ and $P_2 = \int_{-\infty}^{+\infty}|f_2|^2 dx$.

To examine the stability of vector solitons, it is necessary to explore the perturbed solutions for (1) with the form

$$q_{1,2}(x,z) = \exp(i\mu_{1,2}z)\{f_{1,2}(x) + [F_{1,2}(x)\exp(\delta z) + G^*_{1,2}(x)\exp(\delta^* z)]\}, \quad (5)$$

where the superscript '*' represents complex conjugation, $F$ and $G$ the given perturbations with $|F_{1,2}|, |G_{1,2}| \ll |f_{1,2}|$, and $\delta$ a growth rate of perturbations.

Substituting (5) into (1) and then linearizing it, we get a set of equations on the perturbation given as

$$\begin{cases}
\delta F_1 = i\left[\hat{L}_1 F_1 + \sigma\dfrac{(|f_1|^2+|f_2|^2)}{1+s(|f_1|^2+|f_2|^2)}F_1 + \sigma\dfrac{|f_1|^2}{(1+s(|f_1|^2+|f_2|^2))^2}F_1 + \sigma\dfrac{f_1^2}{(1+s(|f_1|^2+|f_2|^2))^2}G_1 + \sigma\dfrac{f_1 f_2^*}{(1+s(|f_1|^2+|f_2|^2))^2}F_2 + \sigma\dfrac{f_1 f_2}{(1+s(|f_1|^2+|f_2|^2))^2}G_2\right] \\
\delta G_1 = i\left[-\sigma\dfrac{f_1^{*2}}{(1+s(|f_1|^2+|f_2|^2))^2}F_1 - \hat{L}_1^* G_1 - \sigma\dfrac{(|f_1|^2+|f_2|^2)}{1+s(|f_1|^2+|f_2|^2)}G_1 - \sigma\dfrac{|f_1|^2}{(1+s(|f_1|^2+|f_2|^2))^2}G_1 - \sigma\dfrac{f_1^* f_2^*}{(1+s(|f_1|^2+|f_2|^2))^2}F_2 - \sigma\dfrac{f_1^* f_2}{(1+s(|f_1|^2+|f_2|^2))^2}G_2\right] \\
\delta F_2 = i\left[\sigma\dfrac{f_2 f_1^*}{(1+s(|f_1|^2+|f_2|^2))^2}F_1 + \sigma\dfrac{f_1 f_2}{(1+s(|f_1|^2+|f_2|^2))^2}G_1 + \hat{L}_2 F_2 + \sigma\dfrac{(|f_1|^2+|f_2|^2)}{1+s(|f_1|^2+|f_2|^2)}F_2 + \sigma\dfrac{|f_2|^2}{(1+s(|f_1|^2+|f_2|^2))^2}F_2 + \sigma\dfrac{f_2^2}{(1+s(|f_1|^2+|f_2|^2))^2}G_2\right] \\
\delta G_2 = i\left[-\sigma\dfrac{f_1^* f_2^*}{(1+s(|f_1|^2+|f_2|^2))^2}F_1 - \sigma\dfrac{f_2^* f_1}{(1+s(|f_1|^2+|f_2|^2))^2}G_1 - \sigma\dfrac{f_2^{*2}}{(1+s(|f_1|^2+|f_2|^2))^2}F_2 - \hat{L}_2^* G_2 - \sigma\dfrac{(|f_1|^2+|f_2|^2)}{1+s(|f_1|^2+|f_2|^2)}G_2 - \sigma\dfrac{|f_2|^2}{(1+s(|f_1|^2+|f_2|^2))^2}G_2\right]
\end{cases}$$
(6)

where, $\hat{L}_1 = \dfrac{1}{2}\dfrac{\partial^2}{\partial x^2} + V + iW - \mu_1$ and $\hat{L}_2 = \dfrac{1}{2}\dfrac{\partial^2}{\partial x^2} + V + iW - \mu_2$.

These equations can be numerically solved to get the perturbation growth rate $\text{Re}(\delta)$ [29]. Solitons are considered linearly stable only if $\max\{\text{Re}(\delta)\} = 0$.

## 3. Numerical studies of Vector Solitons

In this section, we study numerically mixed-gap two-component vector PT solitons in focusing and defocusing saturable media, respectively. The first component is fundamental mode with propagation constant $\mu_1$ and the second component dipole mode with propagation constant $\mu_2$. The existence, stability and evolution properties of vector solitons are discussed in details.

3.1 Vector PT Solitons in the focusing ($\sigma = 1$) saturable nonlinear media

Here, we focus on the properties of mixed-gap fundamental-dipole-coupled vector PT solitons with focusing saturable nonlinearity. We find that the fundamental component belongs to the semi-infinite gap, while the dipole component to the first gap. To explore them, we study the effect of continuously changing the propagation constant $\mu_2$ ($\mu_1$) of the dipole (fundamental) component on the power and stability of such vector solitons. Firstly, we set $\mu_1 = 3$, and calculate the total power (P) and partial powers (P$_1$, P$_2$) of the vector solitons for two different strengths of saturable nonlinearity ($s = 0.1, 0.4$). As is shown in figure 2(a), the total power P of the vector solitons as a function of $\mu_2$ is given for the two different $s$ values. The calculated partial powers versus $\mu_2$ are plotted in figures 2(b) and (c), respectively for $s = 0.1$ and 0.4. For fixed $\mu_1$ and $s$, as can be seen, the dipole component becomes stronger with the increase in $\mu_2$, while the fundamental component becomes gradually weaker. There exist both lower and upper cutoffs of propagation constant $\mu_2$ which determine the existence domain of solitons. In the upper cutoff $\mu_2^{upp}$, fundamental component vanishes, and vector soliton degenerates into scalar dipole mode; in the lower cutoff $\mu_2^{low}$, dipole component vanishes, and one gets scalar fundamental mode. Next, we conduct a numerical stability analysis, with the stable and unstable ranges indicated in figure 2(a) by solid and dashed lines respectively. As is shown, vector solitons are unstable in the high-power region of P.

Next, setting $\mu_2 = 0$, and continuously varying $\mu_1$, we examine the influence of propagation constant of the fundamental component on the properties of vector solitons. As is shown in figure 2(d), the total power curves of vector solitons as a function of $\mu_1$ are plotted for the two different $s$ values. The stable and unstable ranges are also indicated in figure 2(d) by solid and dashed lines respectively. It can be seen that, in this case, solitons are unstable in the low-power region of P. The changes of partial powers of the vector solitons are plotted as a function of $\mu_1$ in figures 2(e) and (f), respectively for $s = 0.1$ and 0.4. It can also be seen that the partial power $P_1$ increases monotonically while the partial power $P_2$ decreases along with the increase in $\mu_1$ for the fixed $\mu_2$ and $s$. The dipole component gradually vanishes when $\mu_1 \to \mu_1^{upp}$, while in the opposite limit $\mu_1 \to \mu_1^{low}$ the fundamental component stop to exist.

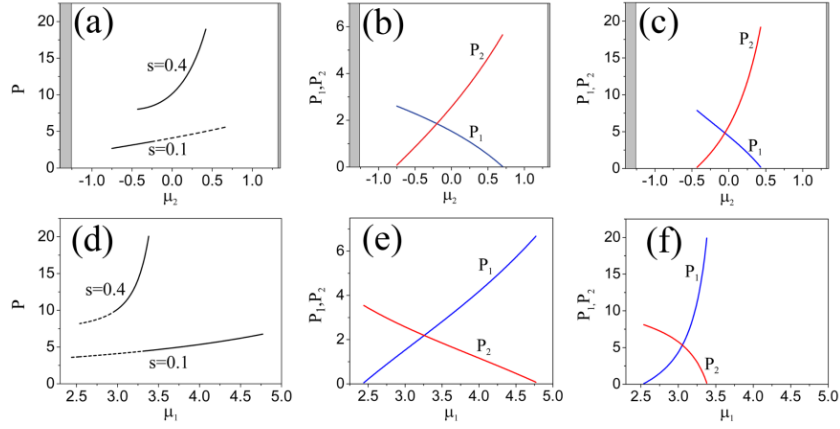

**Figure 2.** (a) The total power of vector solitons versus propagation constant $\mu_2$ (the solid and dashed lines represent the stable and unstable ranges); the partial powers of vector solitons versus propagation constant $\mu_2$ for (b) $s = 0.1$, (c) $s = 0.4$ (the grey region is band); $\mu_1 = 3$. (d) The total power of vector solitons versus propagation constant $\mu_1$ (the solid and dashed lines represent the stable and unstable ranges); the partial powers of vector solitons versus propagation constant $\mu_1$ for (e) $s = 0.1$, (f) $s = 0.4$; $\mu_2 = 0$.

Figures 2(a) and (d) show that the existence domains of vector solitons with $s = 0.1$ is wider than the ones with $s = 0.4$ and that vector solitons are unstable in the high-power region of $P_2$ (not P). In addition, it is worth noting that the increasing rate of soliton power with propagation constant in the deeper saturable nonlinear media ($s = 0.4$) is much higher than in the weak saturable nonlinear media ($s = 0.1$). This can be explained in physics that the deeper saturable effect suppresses the increasing of phase shift induced by the nonlinearity. This property is utilizable for realizing "high power" solitons, even within a limited range of propagation constant (such as finite gaps). More importantly, figure 2(a) clearly shows that for fixed $\mu_1 = 3$, vector solitons given at $s = 0.1$ exist in the range of -0.75 ⩽ $\mu_2$ ⩽ 0.7, but they are stable only within the range of -0.75 ⩽ $\mu_2$ ⩽ -0.22. While for $s = 0.4$, solitons are stable in the whole existence region (-0.43 ⩽ $\mu_2$ ⩽ 0.42). As a result, it is recognizable that some unstable vector solitons in weak saturable nonlinear media can be stabilized by increasing the saturation degree of nonlinear media.

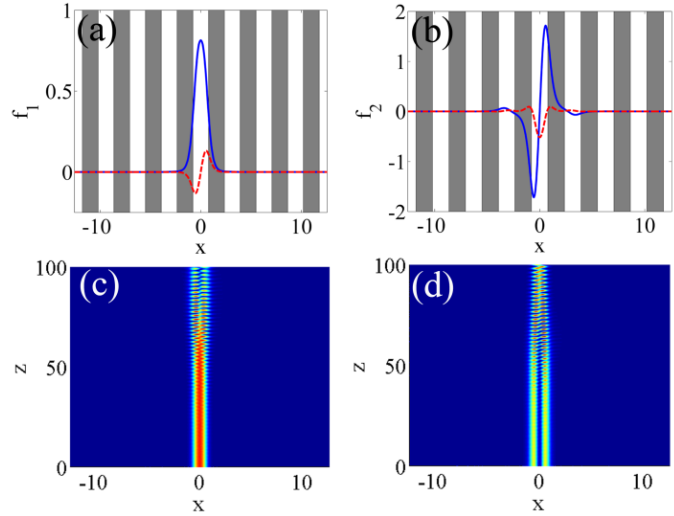

Figure 3. (a) the fundamental component, and (b) the dipole component, (the blue solid and red dashed curves are the real and imaginary parts respectively). (c), (d) Corresponding unstable propagations of the two perturbed components. In all cases, $\mu_1 = 3$, $\mu_2 = 0.4$, $s = 0.1$.

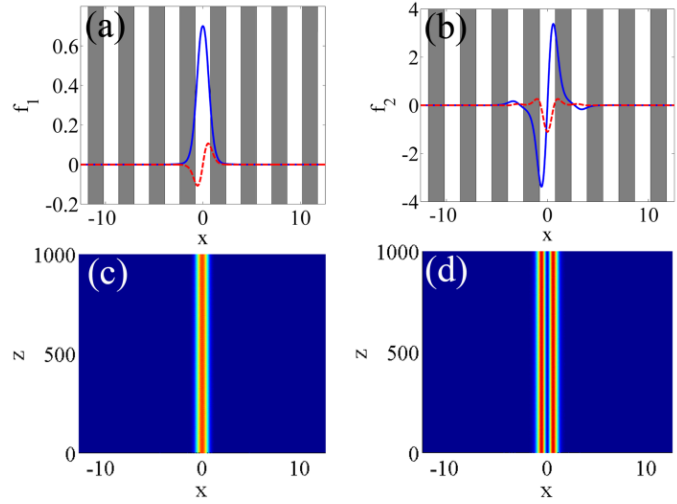

**Figure 4.** (a) The fundamental component, and (b) the dipole component, (the blue solid and red dashed curves are the real and imaginary parts respectively). (c), (d) Corresponding stable propagations of the two perturbed components. In all cases, $\mu_1 = 3$, $\mu_2 = 0.4$, $s = 0.4$.

In our study, two examples with different values of the saturation parameter are selected to illustrate utilizability of the saturation effect of the nonlinear media in suppressing the instability of vector solitons. As can be seen from figure 2(a), for $\mu_2 = -0.21$ and $s = 0.1$, the two-component vector solitons are unstable. Plotted in figures 3(a) and (b) are the soliton profiles of the fundamental mode and dipole mode. The real and imaginary parts of soliton profiles are shown in blue solid and red dashed curves respectively. Figures 3(c) and (d) show the unstable propagations of the two components (5% random noises are added into the initial amplitudes of the two components). When $s = 0.4$, a stable vector soliton can be found, whose profiles of the fundamental component and the dipole component are shown in figures 4(a) and (b) respectively. The two components propagate stably, as shown in figures 4(c) and (d). Propagation simulations of two components verify the linear stability analysis results of vector solitons.

3.2 Vector Solitons in the defocusing ($\sigma = -1$) saturable nonlinear media

Defocusing media (σ = -1) also support mixed-gap PT vector solitons. In this case, the fundamental and dipole components of vector solitons emerge from the first and the second finite gaps respectively. For the case of defocusing saturable nonlinearity, we also analyze the existence and stability properties of solitons with two different saturation degrees ($s=0.1$ and $s = 0.4$). Similarly, the change of P, $P_1$ and $P_2$ are plotted in figure 5 as a function of $\mu_2$ ($\mu_1$) for fixed $\mu_1$ ($\mu_2$). As can be seen from there, with fixed $\mu_1$ there exist upper and lower cutoffs on $\mu_2$, and vice versa. In particular, when other parameters are fixed, with an increase of $\mu_2$ in the fundamental-dipole-coupled solitons shown in figures 5(b) and (c), the dipole component becomes weaker, while its fundamental counterpart becomes stronger. In contrast, the dipole component becomes more pronounced with an increase of $\mu_1$, while the fundamental component would gradually weaken; see figures 5(e) and (f). Furthermore, the results show that similar to the focusing case, (i) as the saturation degree of defocusing nonlinear media is increased, the change of power with propagation constant becomes evidently sharper; and (ii) more powerful saturable effects relate to narrower existence domain of vector solitons.

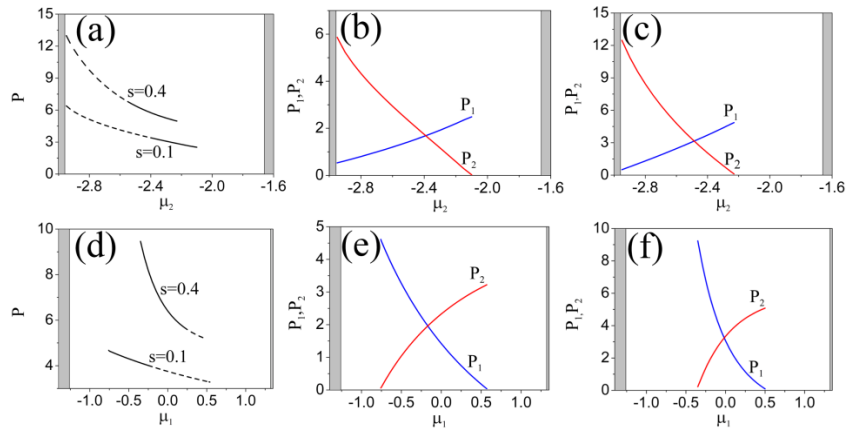

**Figure 5.** (a) The total power of vector solitons versus propagation constant $\mu_2$ (the solid and dashed lines represent the stable and unstable ranges); the partial powers of vector solitons versus propagation constant $\mu_2$ for (b) $s = 0.1$, (c) $s = 0.4$ (the grey region is band); $\mu_1 = 0$. (d) The total power of vector solitons versus propagation

constant $\mu_1$ (the solid and dashed lines represent the stable and unstable ranges); the partial powers of vector solitons versus propagation constant $\mu_1$ for (e) $s = 0.1$, (f) $s = 0.4$; $\mu_2 = -2.5$.

Linear stability analysis is also performed on the vector solitons in the power curves shown in figure 5. In figures 5(a) and (d), there is apparently discernable illustration of the stable and unstable ranges in solid and dashed curves respectively. Obviously, for a part of the existence range of unstable vector solitons with $s = 0.1$, one can find the stable ones when $s = 0.4$. Numerical results indicate that for the defocusing media, deeper saturable nonlinearity is also beneficial to the stability of vector solitons in a limited range. Besides, we find that the stability of such vector solitons is also determined principally by the dipole component instead of the total component, and that relatively low $P_2$ is favorable to the stability of the vector solitons. This suggests that families of unstable scalar high-power fundamental solitons may be stabilized in the vectorial form, due to the coupling to a low-power dipole component. To illustrate this point, we choose a stable (high P but relatively low $P_2$) and an unstable (low P but relatively high $P_2$) examples from figure 5(d). The stable example ($\mu_1 = -0.3$) is shown in figure 6, including solitons incorporating fundamental-mode [figure 6(a)] and dipole-mode [figure 6(b)] components, as well as corresponding propagations [figures 6(c), 6(d)]. By comparing figure 6(a) with figure 6(b), we see that the amplitude of dipole component is much lower than that of fundamental component. The profiles of an unstable vector soliton at $\mu_1 = 0.45$ are plotted in figures 7(a) and (b), which shows that the amplitude of dipole component is higher than that of fundamental one; and the corresponding propagations of two components of the unstable vector soliton are illustrated in figures 7(c) and (d). All the propagation results are in good agreement with our linear stability analyses.

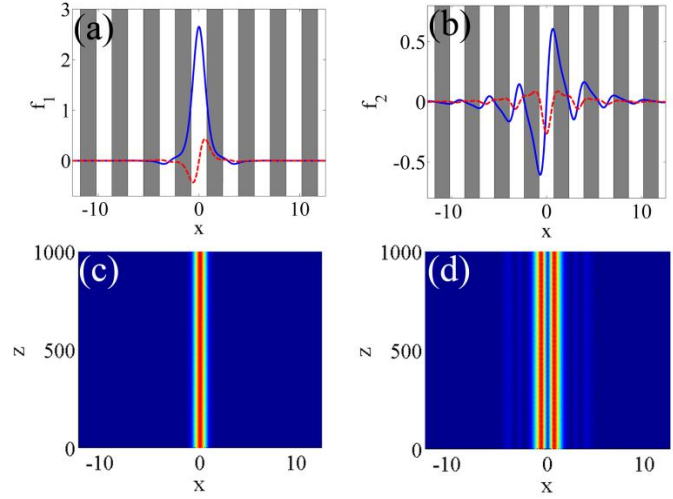

**Figure 6.** (a) The fundamental component, and (b) the dipole component, (the blue solid and red dashed curves are the real and imaginary parts respectively). (c), (d) Corresponding stable propagations of the two perturbed components. In all cases, $\mu_1 = -0.3$, $\mu_2 = -2.5$, $s=0.4$.

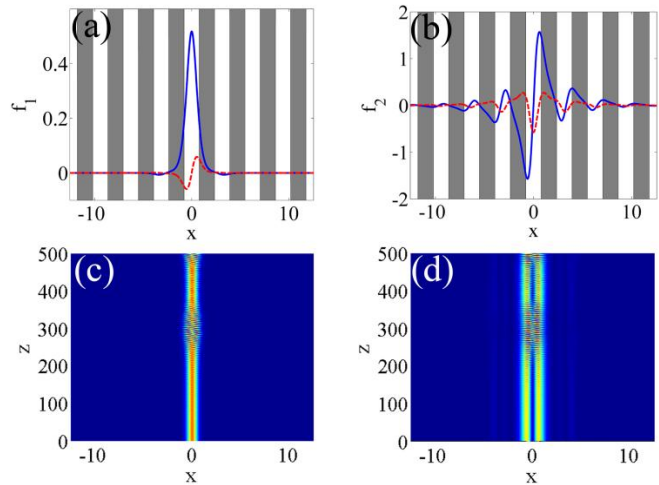

**Figure 7.** (a) the fundamental component, and (b) the dipole component, (the blue solid and red dashed curves are the real and imaginary parts respectively). (c), (d) Corresponding unstable propagations of the two perturbed components. In all cases, $\mu_1 = 0.45$, $\mu_2 = -2.5$, $s = 0.4$.

## 4. Conclusions

In summary, we have reported the existence, stability, and propagation dynamics of new type mixed-gap fundamental-dipole-coupled vector solitons supported by the saturable nonlinear medium and PT optical lattice. Both the focusing and defocusing nonlinearity are analyzed. For the focusing case, the fundamental and dipole components of such vector solitons emerge from the semi-infinite and first gap. For the defocusing case, mixed-gap vector solitons are found too, with the fundamental component from the first gap, while the dipole component from the second gap. Besides, we find that saturable nonlinearity enables to make the soliton power increasing/decreasing sharply with the variation of propagation constant, which is useful for realizing high-power soliton within finite range of propagation constant. What's more, our numerical calculations suggest for both cases, vector solitons with stronger fundamental and weaker dipole components are prone to stability, which offers an additional possibility for experimental observation of solitons made up with high-power fundamental and low-power dipole mode, as high-power solitons may be unstable in scalar case. In particular, we also find that stronger saturable effect is to certain extent helpful for stability of vector solution.

**Acknowledgments**

This work was supported by the National Natural Science Foundation of China under Grants No. 11465009 and No.11165009, the Doctoral Scientific Research Foundation of Jinggangshan University (JZB15002), and the Program for New Century Excellent Talents in University of Ministry of Education of China (NCET-13-0836).